\documentclass[
    aps,
    prd,
    10pt,
    onecolumn,
    superscriptaddress,
    preprintnumbers,
    nofootinbib,
    notitlepage,
    longbibliography,
    floatfix
]{revtex4-1}

\pdfoutput=1 
\usepackage[T1]{fontenc}
\usepackage{graphicx}
\usepackage{paralist}
\usepackage{enumitem}
\usepackage{amsmath}
\usepackage{mathtools}
\usepackage{amsfonts}
\usepackage{amssymb}
\usepackage{bbold}
\usepackage{bm}
\usepackage{multirow}
\usepackage{float}
\usepackage[export]{adjustbox}
\usepackage[dvipsnames]{xcolor}
\usepackage[normalem]{ulem}
\usepackage{orcidlink}
\usepackage{slashed}
\usepackage{hyperref}
\let\oldvec\vec
\renewcommand{\vec}[1]{\bm{\oldvec{#1}}}

\DeclareRobustCommand{\okina}{%
  \raisebox{\dimexpr\fontcharht\font`A-\height}{%
    \scalebox{0.8}{`}%
  }%
}

\newcommand{\dd}{\mathrm{d}}

\usepackage{braket}

\begin{document}
\preprint{DESY-26-123}
\title{Dark Matter-Induced Stellar Oscillations in the de Broglie Regime}

\author{Qiuyue Liang
\orcidlink{0000-0003-1271-6607}} \email{qiuyue.liang@desy.de}
\affiliation{Deutsches Elektronen-Synchrotron DESY, Notkestr. 85, 22607 Hamburg, Germany }
\affiliation{ Kavli IPMU (WPI), UTIAS, University of Tokyo, Kashiwa, 277-8583, Japan}
\author{Jeremy Sakstein \orcidlink{0000-0002-9780-0922}} \email{sakstein@hawaii.edu}
\affiliation{Department of Physics \& Astronomy, University of Hawai\okina i, Watanabe Hall, 2505 Correa Road, Honolulu, HI, 96822, USA}

\begin{abstract}
We investigate stellar oscillations driven by the time-dependent gravitational potential of ultralight dark matter (ULDM) in the de Broglie regime. The response is controlled by the ratio of the ULDM coherence time to the stellar mode damping time. When the field decoheres faster than the damping time, ULDM acts as a stochastic source that excites modes below a characteristic de Broglie frequency, while higher-frequency modes are suppressed. In the opposite, long-coherence regime, individual modes can be resonantly driven. Focusing on solar oscillations, we find a signal far below current sensitivity, and estimate that other stars are similarly unlikely to provide observable signals.
\end{abstract}

\maketitle

\section{Introduction}
\label{sec:intro} 

Ultra-light dark matter (ULDM) refers to a class of models where dark matter (DM) is described by a superposition of classical scalar waves \cite{Hui:2016ltb,Ferreira:2020fam,Hui:2021tkt,Eberhardt:2025caq}.~These models have received interest due to their ability to resolve long-standing small-scale astrophysical challenges to the cold dark matter paradigm such as the cusp-core and missing satellite problems while being theoretically motivated e.g., from QCD axions  \cite{Peccei:1977hh,Weinberg:1977ma,Wilczek:1977pj,Zhitnitsky:1980tq}, the pions of a dark standard model \cite{Maleknejad:2022gyf, Alexander:2023wgk}, the axion-like particles predicted by string theory (see e.g., \cite{Svrcek:2006yi, Arvanitaki:2009fg}), and superfluid DM \cite{Berezhiani:2015bqa,Khoury:2021tvy}. 

The wave-like nature of ULDM predicts novel astrophysical phenomena that provide unique signatures of the theory.~One example is that the gravitational potentials acquire small oscillating components that could manifest in pulsar timing arrays \cite{Khmelnitsky:2013lxt,Porayko:2018sfa,Porayko:2014rfa,EuropeanPulsarTimingArray:2023egv,Kim:2023kyy,Eberhardt:2024ocm,Gan:2025icr,Foster:2026mvs,Lee:2026gzl}, astrometry \cite{Dror:2024con}, binary star evolution \cite{Blas:2016ddr,Blas:2019hxz}, solar system \cite{Frerick:2026kzq}, and gravitational wave interferometers \cite{Aoki:2016kwl,Kim:2023pkx,Gan:2026vph}.

The same time-dependent gravitational field can also drive stellar oscillations. Reference~\cite{Sakstein:2023hvw} developed a framework for calculating the response of stellar oscillation modes to an external gravitational potential and applied it to the zero-momentum or \textit{Compton} mode of the ULDM field (the component with wave number $k=0$), concluding that the resulting signal for the Sun was beyond the reach of solar experiments. Here, we instead apply this framework to the \textit{de Broglie} modes of the field (those with non-zero wave number $k$), which induce stochastic spatial variations of the Newtonian potential. These provide a distinct source of stellar forcing and offer the possibility of an enhanced signal.

Motivated by this, we study the excitation of stellar modes in the de Broglie regime.~We first develop a framework for calculating the amplitude of ULDM-induced stellar oscillations in this regime, identifying two qualitatively different limits.~In the \textit{short-coherence regime}, where the ULDM field decoheres on timescales shorter than the mode damping time, the star acts as a low-pass filter:~modes with frequencies below the characteristic de Broglie frequency are stochastically excited, while higher frequency modes are suppressed.~In the \textit{long-coherence regime}, where the ULDM field is coherent on longer timescales than the mode damping time, the modes are resonantly excited.

The resonant response is exponentially suppressed for realistic stellar oscillation frequencies and halo velocity dispersions, so we focus on the short-coherence regime, where ULDM acts as an additional source of stochastic excitation.~Using the Sun as a case study, we estimate that the signal lies well below the reach of current experimental sensitivity.~We also discuss the prospects for observing ULDM-induced variability in other stellar systems.

This paper is organized as follows.~In section~\ref{sec:ULDM_review}, {we review the nonrelativistic description of ULDM and derive the time-dependent gravitational potential sourced by its de Broglie modes, which provides the stochastic forcing for the stellar oscillations studied in the following sections.}~In section~\ref{sec:Oscillations_review}, we review the excitation of stars by ULDM and extend the formalism to the case of a stochastic ULDM field.~We then develop a framework for calculating the amplitude of ULDM-induced stellar oscillations in section~\ref{sec:framework} and apply it to the Sun in section~\ref{sec:real_stars}.~Finally, we conclude in section~\ref{sec:conclusions}.

\section{Gravitational potential induced by Ultra-Light Dark Matter}
\label{sec:ULDM_review}

Ultra-light dark matter is described by a real scalar $\phi$ with action
\begin{equation}
\label{eq:action}
    S=\int\dd^4x\,\sqrt{-g}\left[\frac{R}{16\pi G}+\frac{1}{2} (\partial_\mu \phi)^2 -\frac{1}{2}   m^2 \phi^2\right],
\end{equation}
where $m$ is the field's mass.~Note that we use the $(+,-,-,-)$ metric signature.~The resulting equation of motion,
\begin{equation}
    \label{eq:EOM}
    (\Box +m^2) \phi = 0,
\end{equation}
has the solution
\begin{equation}
\label{eq:solution}
    \phi(x,t)=\frac{1}{\sqrt{2m}}\left(\psi(\vec{x},t)e^{-imt}+{\textrm{ c.c.}}\right),
\end{equation}
on Minkowski space.~This describes a rapidly oscillating field modulated by a slowly varying complex envelope $\psi(\vec{x},t)$.~

In the non-relativistic limit, where the spatial and temporal variations of $\psi$ are slow compared with the Compton scales, such that
\begin{equation}
\label{eq:NR_def}
|\ddot{\psi}| \ll m |\dot{\psi}|,
\qquad
|\nabla^2\psi| \ll m^2|\psi|,
\end{equation}
Eq.~\eqref{eq:EOM} reduces to
\begin{equation}
\label{eq,schrodinger}
i\dot{\psi}=-\frac{\nabla^2}{2m}\psi.
\end{equation}
To solve this equation, we perform a spatial Fourier transformation, 
\begin{equation}
\label{eq:FourierModes}
\psi(\vec{x},t)=\int\frac{\dd^3\vec{k}}{(2\pi)^3}\psi(\vec{k},t )e^{i \vec{k}\cdot\vec{x}  } ,  
\end{equation}
and find that, 
\begin{eqnarray}
    \psi(\vec{k},t ) = \psi(\vec{k},t = 0 ) e^{- i  \frac{k^2}{2m}t },
\end{eqnarray}
is the solution to the Schr\"{o}dinger equation, Eq.~\eqref{eq,schrodinger}. 
Here, the momentum is related to the velocity through $\vec{k}=m\vec{v}$ and satisfies $k=|\vec{k}|\ll m$, according to Eq.~\eqref{eq:NR_def}. Thus, the ULDM field can be viewed as a superposition of plane waves with wave vector $\vec{k}$ and complex Fourier amplitudes $\psi_i(\vec{k})$. In the following context, we denote the initial amplitude $\psi(\vec{k},t=0)$ by $\psi_i(\vec{k})$.~We define the initial power spectrum of the Fourier modes in this case as
\begin{align}
\braket{\psi_i^* (\vec k)\psi_i (\vec k')  } &= (2\pi)^3 \delta^3 (\vec k -\vec k') P_i (k), \quad\textrm{and} \label{eq:psiPwerSpectrum}\\ \quad \braket{\psi_i^* (\vec k)\psi_i^* (\vec k')  } &= \braket{\psi_i (\vec k)\psi_i (\vec k')  } = 0.
 \end{align}

We can obtain the density from the energy-momentum tensor of the scalar field
\begin{equation}
    \label{eq:EMtensor}
    T_{\mu\nu} =  \partial_\mu \phi\partial_\nu \phi - g_{\mu\nu}\mathcal{L}~\Rightarrow~ \rho (\vec x, t)  = \frac{1}{2} \dot\phi^2 + \frac{1}{2} m^2 \phi^2\ .
\end{equation}
Plugging the non-relativistic expansion, Eq.~\eqref{eq:solution}, into the above expression, we can express the Fourier modes of the energy density as convolutions of the wave field, 
\begin{eqnarray}
   \tilde \rho(\vec{k}, t)=\int_{k^{\prime}}\left(\frac{m}{2}\right)\left(\psi_i^*\left(k^{\prime}-k\right) \psi_i\left({k}^{\prime}\right) e^{i \frac{\left|\vec{k}^{\prime}-\vec{k}\right|^2-\left|{\vec{k}^{\prime}}\right|^2}{2 m} t}+\psi_i\left(k^{\prime}-k\right) \psi_i^*\left(k^{\prime}\right) e^{-i \frac{\left|\vec{k}^{\prime}-\vec{k}\right|^2-\left|\vec{k}^{\prime} \right|^2}{2 m} t}\right),
\end{eqnarray} 
where $\int_k$ is shorthand for $\int \frac{d^3\vec{k}}{(2\pi)^3}$, and we neglect higher-order terms in the momentum expansion, specifically $\mathcal{O}(k^2)$ and higher. We can obtain the power spectrum of $\rho$ at initial time $t=0$
from the convolution of $P_i$,
\begin{eqnarray}
\label{eq,prhoini}
    P_\rho(k, t=0) 
    =\int_{k^{\prime}} m^2 P_i\left(|\vec k^{\prime}-\vec k| \right) P_i\left(k^{\prime}\right)+\mathcal{O}\left(k^4\right)\ . 
\end{eqnarray}
Its auto-correlation can be obtained as 
\begin{eqnarray}
\label{eq,Pi}
\left\langle\psi^*(\vec{k}, t) \psi\left(\vec{k}^{\prime}, t=0\right)\right\rangle=(2 \pi)^3 \delta^3\left(\vec{k}-\vec{k}^{\prime}\right) P_\psi(k, t), P_\psi(k, t)=P_i(k) e^{i \frac{k^2}{2 m} t} ,
\end{eqnarray}
\begin{eqnarray}
\label{eq:P_rho_approximate1}
    P_\rho(k, t)=\int_{k^{\prime}}\left\langle\rho^*(\vec{k}, t) \rho\left(\vec{k}^{\prime}, 0\right)\right\rangle=m^2 \int_{k^{\prime}} P_i\left(|\vec k^{\prime}-\vec k|  \right) P_i\left(k^{\prime}\right) \cos \left(\frac{\vec{k} \cdot \vec{k}^{\prime}}{m} t\right).
\end{eqnarray}

For a given initial power spectrum, one can obtain the auto-correlation of induced gravitational potential through Poisson equation,
\begin{eqnarray}
\label{eq:PoissonPower}
    \nabla^2 \Phi=4 \pi G \rho\Rightarrow \Phi(\vec k, t)=-\frac{4 \pi G}{k^2} \rho(\vec k, t)\Rightarrow P_{\Phi}(k, t)=\frac{16 \pi^2 G^2}{k^4} P_\rho(k, t).
\end{eqnarray}
In the next section, we will discuss the effect of stellar oscillations induced by this oscillating gravitational potential.

\section{Dark-Matter Induced Stellar Oscillations}
\label{sec:Oscillations_review}

Stellar oscillations are linear perturbations of spherically symmetric stars in hydrostatic equilibrium.~In this section, we summarize their relevant features and their excitation by ULDM.~Further details on stellar perturbation theory can be found in references \cite{1980tsp..book.....C,1989nos..book.....U,2001A&A...370..136S,2001A&A...373..916L,2005MNRAS.360..859C,aerts2010asteroseismology,Lopes:2014dba,Lopes:2015pca}, while the forced-mode formalism used below follows reference~\cite{Sakstein:2023hvw}.

\subsection{Stellar Oscillation Theory}
When a star is perturbed, a fluid element initially at (stationary) position $r_0\hat{\vec{r}}$ is displaced by $\vec{\xi}(\vec{r},t)=\vec{\delta r}(\vec{r},t)$ which satisfies $|\vec{\xi}|/R\ll1$ where $R$ is the star's radius.~Linear adiabatic perturbations satisfy 
\begin{align}
\frac{\partial^{2} \vec{\xi}}{\partial t^2} +  \mathcal{L}\vec{\xi} =0,~\label{eq:linearAdiabaticWE}
\end{align}
where $\mathcal{L}$ is a self-adjoint second-order differential operator.~As such, the solutions are characterized by a discrete set of orthogonal oscillation modes $\vec{\xi}_q$ with corresponding eigenfrequencies $\omega_q$ given by 
\begin{equation}
\mathcal{L}\vec{\xi}_q=\omega_q^2\vec{\xi}_q.
\end{equation}
Due to spherical symmetry, there are three quantum numbers $q=\{n\ell m\}$ with $n$ giving the radial order (number of radial nodes) and $\{\ell m\}$ giving the vector spherical harmonic order.~The eigenmodes are orthogonal with respect to the density-weighted inner product, with the normalization defining the mode inertia $I_q$,
\begin{equation}
\int d^3\vec{x}\,\rho_0(\vec{x})\,\vec{\xi}_q^{*}(\vec{x})\cdot\vec{\xi}_{q'}(\vec{x})
=I_q\delta_{qq'},
\end{equation}
where $\rho_0$ is the equilibrium stellar density.~In general, the star oscillates in a superposition of modes:
\begin{equation}
    \vec{\xi}(\vec{r},t)=\sum_{n=0}^\infty\sum_{l=0}^\infty\sum_{m=-l}^l\vec{\xi}_{nlm}(\vec{r})e^{-i\omega_{nlm} t}, 
\end{equation}
with 
\begin{align}
    \vec{\xi}_{q} =\sqrt{4\pi}\left[\xi^r_{nl}(r)\vec{Y}_{lm}(\theta,\phi) + \xi^h_{nl}(r)\vec{\Psi}_{lm}(\theta,\phi) \right],
\label{eq:xi_in_vector_spherical_harmonics}
\end{align}
where $\vec{Y}_{lm}=Y_{lm}\hat{\vec r}$ and $\vec{\Psi}_{lm}=r\vec{\nabla}Y_{lm}$ are the radial and spheroidal vector spherical harmonics;~the toroidal harmonic is absent because the scalar gravitational forcing considered here excites only spheroidal modes.~In general, the perturbations are non-adiabatic and may be sourced, so equation~\eqref{eq:linearAdiabaticWE} is augmented to
\begin{align}
\frac{\partial^{2} \vec{\xi}}{\partial t^2}  + 2\eta \frac{\partial \vec{\xi}}{\partial t} + \mathcal{L}\vec{\xi} =  \vec{\mathcal{{F}}},~\label{eq:fullWE}
\end{align}
where the term proportional to $\partial/\partial t $ represents the damping effects of turbulent viscosity \cite{1977ApJ...212..243G,2005MNRAS.360..859C} parameterized by a frequency-dependent constant $\eta$ \cite{2005MNRAS.360..859C,Chaplin:2008af,2019MNRAS.487..595H}, and the $\vec{\mathcal{{F}}}$ term represents sources of mode forcing. 

\subsection{Stellar Forcing by ULDM}

As noted above, the oscillating gravitational potential sourced by ULDM can drive stellar modes.~Mathematically, the forcing term in Eq.~\eqref{eq:fullWE} acquires a component $\vec{\mathcal{F}}=-\nabla\Phi$.~The  solution of Eq.~\eqref{eq:fullWE} can be expressed as a superposition of modes 
\begin{equation}
\label{eq:velocity_field_expansion}
    \vec{\xi}(\vec{x},t)=\sum_{q}A_q(t)\vec{\xi}_q(\vec{x})e^{-i\omega_qt},
\end{equation}
with amplitude $A_q(t)$.~In the slowly varying amplitude approximation, $|\dot{A}_q|\ll\omega_q|A_q|$ and $|\ddot{A}_q|\ll\omega_q|\dot{A}_q|$, and for a weakly damped oscillator with $\eta\ll\omega_q$, the amplitude satisfies
\begin{equation}
\label{eq:amplitude_equations}
    \frac{\partial A_q}{\partial t}+\eta A_q = \frac{i}{2\omega_q I_q}e^{i\omega_q t}\mathcal{Q}_q(t),
\end{equation}
where  
\begin{equation}
\label{eq:source}
    \mathcal{Q}_q(t)={\sqrt{4\pi}}\int\dd^3\vec{x}\, \Phi(\vec{x},t)\mathfrak{f}_{nl}(r)Y^*_{lm}(\Omega),
\end{equation}
with
\begin{equation}
    \mathfrak{f}_{nl}(r)=\frac{\dd (\rho_0\xi_{nl}^r)}{\dd r}+\frac{2}{r}\rho_0\xi_{nl}^r-\frac{l(l+1)}{r}\rho_0\xi^h_{nl}\ .
\end{equation}
Since, as derived above in equations \eqref{eq:P_rho_approximate1} and \eqref{eq:PoissonPower}, the induced gravitational potential for ULDM in the de Broglie regime is stochastic, the relevant quantity is the auto-correlation of $A_q(t)$.~We now derive this. 

\subsection{Amplitude of ULDM-Induced Oscillations}

The general solution of equation \eqref{eq:amplitude_equations} is
\begin{equation}
    A_q(t)=\frac{i}{2\omega_qI_q}\int_{-\infty}^t\dd t' e^{i\omega_qt'}e^{\eta(t'-t)}\mathcal{Q}_q(t'),
\end{equation}
where the lower limit is taken as $t'\rightarrow-\infty$ to remove the transient solution.~Changing variables to $\tau=t-t'$, we find
\begin{equation}
\label{eq:amp_tau}
    A_q(t)=\frac{i}{2\omega_qI_q}\int_0^{\infty}\dd \tau \,e^{i\omega_q(t-\tau)}e^{-\eta\tau}\mathcal{Q}_q(t-\tau),
\end{equation}
from which it follows that
\begin{eqnarray}
\label{eq:fullAmp}
    \langle|A_q(t)|^2\rangle 
    &=& \frac{1}{4\omega_q^2I_q^2}\int_0^{\infty}\dd \tau\int_0^{\infty}\dd \tau' e^{-(\eta+i\omega_q)\tau}e^{-(\eta-i\omega_q)\tau'}\langle\mathcal{Q}_q(t-\tau)\mathcal{Q}^*_q(t-\tau')\rangle \ ,
\end{eqnarray}
 and 
 \begin{eqnarray}
 \label{eq,autoQq}
     \langle\mathcal{Q}_q(t-\tau)\mathcal{Q}^*_q(t-\tau')\rangle &=&  4\pi \int d^3 \vec x_{1,2} \mathfrak{f}_{nl}(r_1)Y^*_{lm}(\Omega_1)\mathfrak{f}_{nl}(r_2)Y_{lm}(\Omega_2) \braket{\Phi(\vec x_1, t-\tau) \Phi^*(\vec x_2, t-\tau') } \nonumber\\
     &=& 4\pi \int d^3 \vec x_{1,2} \frac{d^3 \vec k}{(2\pi)^3} \mathfrak{f}_{nl}(r_1)\mathfrak{f}_{nl}(r_2)Y^*_{lm}(\Omega_1) Y_{lm}(\Omega_2) P_\Phi(k, \tau'-\tau) e^{i \vec k\cdot(\vec x_1 -\vec x_2)} \ .
 \end{eqnarray}
This means we need to calculate the autocorrelation spectrum for $\Phi$ in Eq.~\eqref{eq,autoQq}.~To do so, we need a model of ULDM as further discussed in the next section. 
 
\section{Stellar Excitation in the de Broglie Regime}
\label{sec:framework}

We now apply the theory introduced in Secs.~\ref{sec:ULDM_review} and \ref{sec:Oscillations_review} to calculate the amplitude of ULDM-induced stellar oscillations in the de Broglie regime.~Since this is inherently stochastic, it is more instructive to calculate $\langle|A_q (t)|^2\rangle$.~This section is therefore organized as follows.~ We first specialize the general result of Section~\ref{sec:ULDM_review} to a virialized halo with a Maxwell--Boltzmann velocity distribution.~Two regimes naturally arise:~the short-coherence regime where the ULDM decoheres before the mode decays, and the long-coherence regime where the field is coherent over many cycles.~We calculate the amplitude in both limits, which we subsequently use in later sections to guide observational strategies.

\subsection{Virialized ULDM Halos}

In virialized halos, the Fourier modes can be approximated as following a Maxwell-Boltzmann (MB) velocity distribution \cite{Eberhardt:2024ocm}:
\begin{equation}
    f(v)=\frac{1}{({2\pi\sigma^2})^{3/2}}e^{-\frac{v^2}{2\sigma^2}}\ ,
\end{equation}
where $\sigma$ is the local DM velocity dispersion and the normalization factor is chosen such that $\int \dd^3 \vec{v} f(v) =1$.~ The initial power spectrum defined in Eq.\eqref{eq:psiPwerSpectrum} can then be expressed as
\begin{eqnarray}
\label{eq:powerSpectrumFourierMaxwellBoltzmann}
    P_i (k)  = (2\pi)^3\frac{\bar\rho}{m}\frac{1}{({2\pi m^2 \sigma^2})^{3/2}}e^{-\frac{k^2}{2m^2\sigma^2}} \ , 
\end{eqnarray}
where $\vec k = m \vec v$, and the normalization factor is chosen such that the stochastic amplitudes $m\langle |\psi|^2\rangle=\bar \rho$, where $\bar\rho$ is the average density of the local ULDM.

To evaluate the auto-correlation of the energy density, $P_\rho$, in Eq.\eqref{eq:P_rho_approximate1} analytically, we make a series of approximations to compute the integral. First, note that the  Maxwell-Boltzmann speed distribution, $v^2f(v)$, is peaked near $v\sim\sqrt{2}\sigma$ so the integral has the most support when $k',\,(|\vec{k}-\vec{k}'|)\sim m\sigma\ll m$.~Second, we can approximate the MB distribution and the initial power spectrum as
\begin{equation}
    k^2f(v)\approx \frac{m^3}{4\pi }\delta(k-k_0)\ ,\ k^2 P_i(k) \approx (2\pi)^3 \frac{\bar\rho}{m} \frac{\delta(k-k_0)}{4\pi}\ ,
\end{equation}
where $k_0=\sqrt{2}\,m\sigma$ is the peak of the MB speed distribution $4\pi v^2 f(v)$.~With these approximations, equation \eqref{eq:P_rho_approximate1} becomes  
\begin{align}
    P_\rho (k,t) &=\frac{(2\pi)^3\bar \rho^2}{4\pi m^3}\int{\dd k' \dd^2\Omega_{k'}}\, f\left(\frac{|\vec{k}-\vec{k}'|}{m}\right)\delta(k'-k_0)\cos\left(\frac{{\vec{k}}\cdot{\vec{k}}'}{m}t\right)\nonumber\\
    &=\frac{(2\pi)^4\bar \rho^2}{4\pi m^3(2\pi\sigma^2)^{3/2}}e^{-\frac{k^2+k_0^2}{2m^2\sigma^2}}\int_{-1}^1{ \dd\mu}\, e^{\frac{kk_0}{m^2\sigma^2}\mu}\cos\left(\frac{kk_0\mu}{m}t\right)\nonumber\\
    &=\frac{(2\pi)^{3/2}\bar \rho^2}{2m^3\sigma^{3}}\frac{k_0e^{-\frac{k^2+k_0^2}{2m^2\sigma^2}}}{k}\mathcal{I}(k,t)\ ,
\end{align}
where $\mu = \vec k\cdot\vec k'/ (|\vec k||\vec k'|)$, and   
\begin{align}
  \mathcal{I}(k,t) =\frac{\frac{t}{\tau_{\rm dB}}\cosh\beta\sin(\alpha t)+\sinh\beta\cos(\alpha t)}{1+t^2/\tau_{\rm dB}^2}\ , 
\end{align}
with 
\begin{equation}
    \tau_{\rm dB}= \frac{1}{m\sigma^2},\quad \alpha= \sqrt{2}\,k\sigma,\quad\textrm{and}\quad \beta=\frac{\sqrt{2}\,k}{m\sigma}.
\end{equation}
Note that $\tau_{\rm dB}=(m\sigma^2)^{-1}$ is the de Broglie coherence time, over which ULDM modes with different halo velocities dephase.

Finally, the power spectrum for the gravitational potential following from equation \eqref{eq:PoissonPower} is:
\begin{equation}
P_\Phi(k,t)=\frac{8\sqrt{2}\pi^2 (2\pi)^{3/2} G^2\bar \rho^2}{m^2\sigma^{2}}\frac{e^{-\frac{k^2+k_0^2}{2m^2\sigma^2}}}{k^5}\mathcal{I}(k,t)\ .
\end{equation}

To make further progress we need to  simplify $\mathcal{I}(k,t)$.~Examining Eq.~\eqref{eq:amp_tau}, we identify two contrasting limits:~the \textit{short coherence regime} $t\sim\eta^{-1}\gg\tau_{\rm dB}$ i.e., the field decoheres before the mode damps, and the \textit{long coherence regime} $t\sim\eta^{-1}\ll\tau_{\rm dB}$ i.e., the field is coherent over many oscillation cycles.~We now investigate each scenario in turn.

\subsubsection{Short Coherence Regime}

When $t\sim\eta^{-1}\gg\tau_{\rm dB}$ we can take 
\begin{equation}
    \mathcal{I}(k,t)=\frac{\tau_{\rm dB}}{t}\cosh\beta\sin(\alpha t).
\end{equation}
Using this in equations  \eqref{eq:fullAmp} and \eqref{eq,autoQq} we find
\begin{align}
\nonumber
\langle|A_q|^2\rangle&=\frac{8\sqrt{2}\pi^3  G^2\bar \rho^2\tau_{\rm dB}}{ (2\pi)^{3/2}I_q^2\omega_q^2m^2\sigma^{2}}\int\dd^3\vec{k}\,\dd^3\vec{x}\,\dd^3\vec{x}'\,e^{i\vec{k}\cdot{(\vec{x}-\vec{x}')}}\mathfrak{f}_{nl}(r)\mathfrak{f}_{nl}(r')Y^*_{lm}(\hat r)Y_{lm}(\hat r')\frac{\cosh(\beta)e^{-\frac{k^2+k_0^2}{2m^2\sigma^2}}}{k^5}\times\\&\int_0^{\infty}\dd \tau\int_0^{\infty}\dd \tau' e^{-(\eta+i\omega_q)\tau}e^{-(\eta-i\omega_q)\tau'}
    \frac{\sin[\alpha (\tau'-\tau)]}{\tau'-\tau}.\label{eq:dtau_ints}
\end{align}
We can evaluate the $\tau$ and $\tau'$ integrals in \eqref{eq:dtau_ints} as
\begin{align}
    \int_0^{\infty}\dd \tau\int_0^{\infty}\dd \tau' e^{-(\eta+i\omega_q)\tau}e^{-(\eta-i\omega_q)\tau'}
    \frac{\sin[\alpha (\tau'-\tau)]}{\tau'-\tau}&=\frac{1}{4\eta}\left[\arctan\left(\frac{\alpha+\omega_q}{\eta}\right)+\arctan\left(\frac{\alpha-\omega_q}{\eta}\right)\right]\nonumber\\&\approx\frac{\pi}{4\eta}\Theta(\alpha-\omega_q),
\end{align}
where the last line took the limit $\eta\ll |\alpha\pm\omega_q|$.~This tells us that, in the short coherence regime, the system acts like a low pass filter because only modes with $\omega_q<\sqrt{2}k\sigma$ are excited.~Putting this result into eq.~\eqref{eq:dtau_ints} we find
\begin{align}
\label{eq:dtau_ints2}
\nonumber
\langle|A_q|^2\rangle&=\frac{2\sqrt{2}\pi^4  G^2\bar \rho^2\tau_{\rm dB}}{ (2\pi)^{3/2}I_q^2\omega_q^2m^2\sigma^{2}\eta}\int\dd^3\vec{k}\,\dd^3\vec{x}\,\dd^3\vec{x}'\,e^{i\vec{k}\cdot{(\vec{x}-\vec{x}')}}\mathfrak{f}_{nl}(r)\mathfrak{f}_{nl}(r')Y^*_{lm}(\hat r)Y_{lm}(\hat r') \\&\qquad \qquad \times\frac{\cosh(\beta)e^{-\frac{k^2+k_0^2}{2m^2\sigma^2}}}{k^5}\Theta(\alpha-\omega_q).
\end{align}
We can make progress on the angular integrals by expanding the exponentials in plane-waves using
\begin{equation}   e^{i\vec{k}\cdot\vec{x}}=4\pi\sum_{l,m}i^lj_l(kr)Y_{lm}(\hat{r})Y^*_{lm}(\hat{k}),
\end{equation}
where $j_l$ are spherical Bessel functions to find
\begin{align}
\label{eq:dtau_ints3}
\langle|A_q|^2\rangle&=\frac{32\sqrt{2}\pi^6  G^2\bar \rho^2\tau_{\rm dB}}{ (2\pi)^{3/2}I_q^2\omega_q^2m^2\sigma^{2}\eta}\int_{k_q}^\infty\dd{k}\,|\mathcal{W}_{nl}(k)|^2\frac{\cosh(\beta)e^{-\frac{k^2+k_0^2}{2m^2\sigma^2}}}{k^3},
\end{align}
with
\begin{equation}
\label{eq,modefn}
    \mathcal{W}_{nl}(k)=\int\dd r\, r^2j_l(kr)\mathfrak{f}_{nl}(r),
\end{equation}
and $k_q=\omega_q/\sqrt{2}\sigma$.~In deriving Eq.~\eqref{eq:dtau_ints3}, we have used the orthonormality relation
\begin{equation}
\int d^2\Omega\,Y_{\ell m}(\Omega)Y^*_{\ell' m'}(\Omega)
=\delta_{\ell\ell'}\delta_{mm'}.
\end{equation}
To make further progress we expand the $\cosh$ function as the sum of two exponentials and complete the square in the integrand to find 
\begin{equation}
    e^{-\frac{k^2+k_0^2}{2m^2\sigma^2}}\cosh\left(\frac{\sqrt{2}k}{m\sigma}\right)=\frac12\left[e^{-\frac{(k-k_0)^2}{2m^2\sigma^2}}+e^{-\frac{(k+k_0)^2}{2m^2\sigma^2}}\right].
\end{equation}
Now the second term is an exponentially decaying tail while the first is a Gaussian peaked at $k=k_0$ with width $\sim m\sigma\sim k_0$.~We therefore neglect the second term.~If $k_0<k_q$ then the peak is not sampled in \eqref{eq:dtau_ints3} while in the opposite case the peak contributes the most in the integrand. In this case we can approximate the Gaussian as a delta function:
\begin{equation}
\label{eq:bad_delta_approx}
    \frac12e^{-\frac{(k-k_0)^2}{2m^2\sigma^2}}\approx \frac12\sqrt{2\pi}m\sigma\delta(k-k_0),
\end{equation}
so that \eqref{eq:dtau_ints3} gives:
\begin{align}
\langle|A_q|^2\rangle&=\frac{4\pi^5  G^2\bar\rho^2}{ (2\pi)^{3/2}I_q^2\omega_q^2m^5\sigma^{6}\eta}|\mathcal{W}_{nl}(k_0)|^2 \ , \quad k_0>k_q.\label{eq:dtau_ints4}
\end{align}
This expression is only valid for $k_0>k_q$ i.e., $\omega_q<2m\sigma^2$.~If this is not satisfied then the peak is not sampled by eq.~\eqref{eq:dtau_ints3} and there is an exponential suppression.

\subsubsection{Long Coherence Regime}

When $t\sim\eta^{-1}\ll\tau_{\rm dB}$ we can take 
\begin{align}
  \mathcal{I}(k,t) \approx \sinh\beta\cos(\alpha t). 
\end{align}
Following the same steps as above, we find 
\begin{align}
\nonumber\langle|A_q|^2\rangle&=\frac{8\sqrt{2}\pi^3  G^2\bar \rho^2}{ (2\pi)^{3/2}I_q^2\omega_q^2m^2\sigma^{2}}\int\dd^3\vec{k}\,\dd^3\vec{x}\,\dd^3\vec{x}'\,e^{i\vec{k}\cdot{(\vec{x}-\vec{x}')}}\mathfrak{f}_{nl}(r)\mathfrak{f}_{nl}(r')Y^*_{lm}(\theta,\phi)Y_{lm}(\theta',\phi')\frac{\sinh(\beta)e^{-\frac{k^2+k_0^2}{2m^2\sigma^2}}}{k^5}\times\\&\int_0^{\infty}\dd \tau\int_0^{\infty}\dd \tau' e^{-(\eta+i\omega_q)\tau}e^{-(\eta-i\omega_q)\tau'}
    \cos[\alpha (\tau'-\tau)].\label{eq:dtau_ints_long}
\end{align}
We can evaluate the $\tau$ and $\tau'$ integrals in \eqref{eq:dtau_ints_long} as
\begin{align}
\label{eq,timeintegration}
\nonumber
    \int_0^{\infty}\dd \tau\int_0^{\infty}\dd \tau' e^{-(\eta+i\omega_q)\tau}e^{-(\eta-i\omega_q)\tau'}
    \cos[\alpha (\tau'-\tau)]&=\frac{1}{2}\left[\frac{1}{\eta^2+(\omega_q+\alpha)^2}+\frac{1}{\eta^2+(\omega_q-\alpha)^2}\right]\\&\approx\frac{\pi}{2\eta}\delta(\omega_q-\alpha),
\end{align}
where we approximated the resonance in the second line by a delta function.~Using this in \eqref{eq:dtau_ints_long} we find
\begin{align} 
\langle|A_q|^2\rangle=\frac{128\sqrt{2}\pi^6  G^2\bar \rho^2}{ (2\pi)^{3/2}I_q^2\omega_q^5m^2\eta}|\mathcal{W}_{nl}(k_q)|^2e^{-\frac{\omega_q^2\tau_{\rm dB}^2}{4}-1}\sinh\left(\omega_q\tau_{\rm dB}\right), \label{eq:dtau_ints5_long}
\end{align}
where $k_q=\omega_q/\sqrt{2}\sigma$.~There are two regimes of interest. When $\omega_q\tau_{\rm dB}\gg1$ the amplitude is exponentially suppressed.~Physically, the ULDM $k$-mode needed to excite the stellar mode, $k_q$, is in the exponential tail of the Maxwell-Boltzmann distribution.~In the opposite limit, $\omega_q\tau_{\rm dB}\ll1$, we have  
\begin{align}
\label{eq:dtau_ints6_long}
\langle|A_q|^2\rangle&=\frac{128\sqrt{2}\pi^6  G^2\bar \rho^2\tau_{\rm dB}}{ (2\pi)^{3/2}e I_q^2\omega_q^4m^2\eta}|\mathcal{W}_{nl}(k_q)|^2\ .
\end{align}

\section{Observational Implications}
\label{sec:real_stars}

{In this section, we discuss the observational implications of the formalism developed above. We first identify the conditions under which ULDM can efficiently excite stellar oscillations, and then apply the results to the Sun and other stellar systems.}

\subsection{Conditions for Observable Excitations}

The calculation in the previous section identified two regimes of ULDM-induced stellar excitation:
\begin{compactitem}
    \item[\textbf{Short-coherence regime ($\eta\tau_{\rm dB}\ll1$):}] The field is incoherent on timescales of order the mode life. The ULDM acts as a source of stochastic excitation with amplitude given in Eq.~\eqref{eq:dtau_ints4}.~Only modes with $\omega_q<2m\sigma^2$ are excited.
    \item[\textbf{Long-coherence regime ($\eta\tau_{\rm dB}\gg1$):}] The field is coherent on a timescale longer than the mode life. The field resonantly excites individual modes, however the amplitude is exponentially suppressed unless $\omega_q\tau_{\rm dB}\ll1$.
\end{compactitem}
These considerations provide observational guidance for searching for ULDM-induced stellar oscillations.

Note that
\begin{equation}
    \frac{\eta\tau_{\rm dB}}{\omega_q\tau_{\rm dB}}=\frac{\eta}{\omega_q}
\end{equation}
so to be in the long-coherence regime and avoid the exponential suppression we need $\eta\gg\omega_q$.~This implies that the object is in the strongly damped regime, which violates the assumption of weak damping used in our derivation, so a separate framework is needed to assess this scenario.~Since stellar oscillators occupy the weak damping regime, the considerations above imply that the short-coherence regime is most relevant.

\subsection{Application to the Sun}

The Sun's modes have cyclic frequencies of order $\nu_q=\omega_q/(2\pi)\sim3000\,\mu{\rm Hz}$ and are in the short-coherence regime for $m\gtrsim10^{-16}$ eV.~The relevant observable is the root-mean-square (RMS) surface velocity, given by (see \cite{Sakstein:2023hvw} equation~(24)) 
\begin{align}
\label{eq:Alm_short_coherence_numbers}
    V_{\rm rms}&=\omega_q|\vec{\xi}_q|_{r=R}\sqrt{\langle|A_q|^2\rangle} \nonumber\\&= 2.9\times10^{-19}C_{nl}\left(\frac{\rho}{0.42\,{\rm GeV/cm}^3}\right)\left(\frac{2\times10^{-11} {\rm eV}}{m}\right)^{\frac52}\left(\frac{10^{-7} {\rm s}^{-1}}{\eta}\right)^{\frac12}\left(\frac{220\,{\rm km/s}}{\sigma}\right)^{3}\,{\rm cm/s},
\end{align}
where
\begin{equation}
    C_{nl}\equiv \left(\frac{\xi_q \mathcal{W}_{\rm nl}}{I_q}\right)R_\odot.
\end{equation}
The fiducial values correspond to the DM density in the solar neighborhood ($0.42$ GeV/cm$^3$), the typical solar mode damping rate ($\eta\sim10^{-7}$s$^{-1}$) and the Milky Way velocity dispersion at the solar location ($\sigma\sim220$ km/s).~In addition, Eq.~(\ref{eq:Alm_short_coherence_numbers}) applies only when $\omega_q<2m\sigma^2$;~for $\nu_q\sim3000\,\mu{\rm Hz}$ and $\sigma=220$ km/s, this requires $m\gtrsim1.1\times10^{-11}$ eV, motivating our benchmark $m=2\times10^{-11}$ eV.~On dimensional grounds, we anticipate that $C_{nl}\sim\mathcal{O}(1)$.~We verified this by applying GYRE \cite{2013MNRAS.435.3406T} to the MESA \cite{Paxton:2010ji,Paxton:2013pj,Paxton:2015jva,Paxton:2017eie,Paxton:2019lxx,MESA:2022zpy} solar model of reference \cite{Sakstein:2023hvw}, finding $C_{nl}\sim\mathcal{O}(1$--$10)$ for representative solar modes.~Typical solar RMS velocities are $\mathcal{O}(1\,\textrm{cm/s})$ \cite{GoudekPhD, Libbrecht1998} so we conclude that ULDM halo-induced solar oscillations in the de Broglie regime are unlikely to be observed.  

\subsection{Other Stellar Types}

The solar calculation above naturally raises the question of whether ULDM-induced oscillations may be more readily observed in other types of stars.~For distant stars, the observable is photometric variability rather than direct radial-velocity measurements.~We can estimate this by perturbing the Stefan--Boltzmann law, $L=4\pi R^2\sigma_{\rm SB} T^4$, and assuming the motion is adiabatic to find the fractional change in the luminosity:
\begin{equation}
    \frac{\Delta L}{L}=2\frac{\Delta R}{R}+4\frac{\Delta T}{T}=2\frac{\Delta R}{R}+4\left(\Gamma_3-1\right)\frac{\Delta \rho}{\rho}\ ,
\end{equation}
where $\Gamma_3$ is the third adiabatic index.~Using the relation \cite{1980tsp..book.....C,aerts2010asteroseismology}
\begin{equation}
    \frac{\Delta \rho}{\rho}=-\nabla\cdot\vec{\xi}\sim -\frac{\xi(R)}{R}\ ,
\end{equation}
one finds the surface luminosity variation to be
\begin{equation}
     \frac{\Delta L}{L}\sim\left[2-4\left(\Gamma_3-1\right)\right]\frac{\xi(R)}{R}\ .
\end{equation}
The RMS variation in the $q$th mode is then
\begin{equation}
     \left.\frac{\Delta L}{L}\right\vert_{\rm rms}\sim\left[2-4\left(\Gamma_3-1\right)\right]\frac{V_{\rm rms}}{\omega_qR}\ .
\end{equation}
This scales as
\begin{equation}
     \left.\frac{\Delta L}{L}\right\vert_{\rm rms}\sim 6.7\times10^{-28}\mathcal{C}_{nl}\left(\frac{\rho}{0.42\,{\rm GeV/cm}^3}\right)\left(\frac{2\times10^{-11} {\rm eV}}{m}\right)^{\frac52}\left(\frac{10^{-7} {\rm s}^{-1}}{\eta}\right)^{\frac12}\left(\frac{220\,{\rm km/s}}{\sigma}\right)^{3}\left(\frac{1000 \mu{\rm Hz}}{\nu_q}\right)\left(\frac{R_\odot}{R}\right)^2,\label{eq:Lrms}
\end{equation}
where $\nu_q=\omega_q/2\pi$ is the mode frequency and $\mathcal{C}_{nl}$ is defined via
\begin{equation}
 \mathcal{C}_{nl}\equiv   \left(\frac{\xi_q \mathcal{W}_{\rm nl}}{I_q}\right)R.
\end{equation}
{Equation~\eqref{eq:Lrms} suggests that ULDM-induced stellar oscillations are unlikely to be observable for a smooth, virialized ULDM halo with a Maxwell--Boltzmann velocity distribution. For characteristic local dark-matter densities and velocity dispersions, neither weaker damping nor the smaller velocity dispersions found in systems such as dwarf spheroidal galaxies, $\sigma\sim10~{\rm km\,s^{-1}}$, appear sufficient to overcome the strong suppression of the signal.} 

\section{Conclusions}
\label{sec:conclusions}

In this work, we studied the excitation of stellar oscillations by ultralight dark matter in the de Broglie regime. In this scenario, finite-momentum ULDM modes source a stochastic Newtonian potential that drives stellar oscillations. We developed a formalism for calculating the resulting mode amplitudes in terms of the stellar properties and environment.

The response is controlled by the competition between the ULDM de Broglie coherence time and the mode damping time. When the field decoheres before the mode damps, ULDM acts as a stochastic source of excitation. In this short-coherence regime, the stellar response behaves as a low-pass filter, so that only sufficiently low-frequency modes are efficiently excited. In the opposite long-coherence regime, individual modes can be resonantly driven. However, for ordinary underdamped stellar oscillations, the resonance lies in the exponentially suppressed tail of the halo velocity distribution.

Applying our results to solar oscillations, we find an induced surface velocity far below current experimental sensitivity. We further showed that recasting the signal in terms of a photometric luminosity variation does not generically evade this suppression. 

The methodology developed in this paper can also be adapted to the mini-cluster scenario \cite{Fairbairn:2017sil, Hardy:2016mns, Kirkpatrick:2020fwd}, where the local energy density can be significantly enhanced compared to the ULDM halo case, potentially leading to larger solar oscillation signals. We leave a detailed discussion of this scenario to future work.

\section*{Acknowledgments}
We thank M.~Amin, A.~Eberhardt, E.~Ferreira, S.~Oishi, and Y.~Wang for helpful discussions.~J.S.~thanks the Kavli Institute for the Physics and Mathematics of the Universe (IPMU), for its hospitality during a visit in which this work was initiated. Q.L. is supported by the World Premier International Research Center Initiative (WPI), MEXT, Japan (Kavli IPMU) and Deutsche Forschungsgemeinschaft under Germany’s Excellence Strategy - EXC
2121 Quantum Universe - 390833306.

\appendix

\bibliography{ref_ULDM}
\end{document}